\documentclass[11pt]{article}
\usepackage{moriond,epsfig}
\usepackage{graphicx}

\def\be{\begin{equation}}
\def\ee{\end{equation}}
\def\bea{\begin{eqnarray}}
\def\eea{\end{eqnarray}}

\begin{document}
\vspace*{4cm}
\title{HEAVY-ION PHYSICS WITH CMS}

\author{A. IORDANOVA (for the CMS Collaboration)}

\address{Department of Physics, 845 W. Taylor St, M/C 273\\
Chicago, IL, 60607, USA}

\maketitle\abstracts{This article presents a brief overview of
the CMS experiment capabilities to study the hot and dense
matter created in relativistic heavy-ion collisions.  The CERN Large Hadron
Collider will provide collisions of Pb nuclei at 5.5~TeV per nucleon. The CMS
heavy ion group has developed a plethora of physics analyses
addressing many important aspects of heavy-ion physics in
preparation for a competitive and successful program.}

\section{Introduction}

The CMS heavy ion program will study the QCD matter created
in Pb+Pb collisions at the Large Hadron Collider (LHC), at
CERN.  The Pb nuclei will be collided at an energy of 5.5~TeV
per nucleon,
the highest ever reached by a particle accelerator. The
energy density will surely exceed that currently accessible
at the Relativistic Heavy Ion Collider (RHIC), allowing us to
explore a novel QCD regime, where semi-hard and hard processes
will be dominant. The CMS heavy ion program will study all
aspects of particle production from soft physics at relatively
low momenta, to hard probes and ultra-peripheral collisions.
The broad range of physics topics and future analysis
possibilities are summarised in the CMS Heavy Ion Technical
Design Report on high-density QCD with heavy-ions~\cite{cite:pTDR}.
Here, only a very brief summary
of some of the planned physics analyses will be discussed.

The CMS detector is well equipped to carry out a successful heavy
ion program.  The detector has a layered design with full azimuthal
coverage over a wide pseudorapidity range.  The main systems --- the
Silicon tracker, the electromagnetic (ECAL) and hadronic (HCAL)
calorimeters, the muon systems and the solenoidal magnet --- are
detailed in the Technical Design Report~\cite{cite:TDR}.

The global event characterisation measurements will extensively
rely on information from the Silicon tracker (both the pixel and
strip layers), which covers the pseudorapidity window $|\eta|<2.5$.
The tracker has a momentum resolution of better than 2\% for tracks
with $p_{_{T}}<100$~GeV/$c$ ($|\eta|<0.5$) thanks to the powerful
4~T magnetic field.  This detector will provide both charged particle
tracking and vertex reconstruction.  Given the high granularity
of the pixels, the occupancy for the inner pixel layers is expected
not to exceed the 3\% level, even in the heavy ion environment.

The CMS calorimetry and muon identification systems will provide
information for the measurement of specific probes,
such as identified $\gamma$ and jets.  CMS has a
very extensive pseudorapidity coverage of the electromagnetic
($|\eta|$$<$$3$) and hadronic ($|\eta|$$<$$5.2$) calorimeters,
useful, in particular, for the measurement of jets.  This
coverage is further extended by the Castor (5$<$$|\eta|$$<$6.6)
and ZDC ($|\eta|$$>$8.1, for neutrals) forward calorimeters.

The mid-rapidity muon chambers, interspersed in the return yoke
magnetic field (2~T) will provide a very fast and precise
measurement of muon position and (when coupled with the tracker) momentum.
This clean sample of muons will be used to reconstruct quarkonia 
(J/$\psi$ and $\Upsilon$ family) via the di-muon channel.

CMS has an elaborate triggering system.  The combination of fast 
Level-1 hardware systems and powerful computational High Level
Trigger will be crucial for a successful heavy ion program. 

\section{Soft physics}

The measurement of identified particles is essential to fully
characterise the global properties of the created medium:
the initial energy density, the freeze-out properties of the
system, and its hadro-chemistry.  For this purpose, tracking 
algorithms have been developed for low momentum
(p$_{\rm _{T}}$$<$1.5~GeV/$c$) charged particle track
reconstruction, using the pixel detector.  In this kinematic
range, the ionisation energy loss in the pixel layers is 
exploited to identify $\pi^{\pm}, K^{\pm}$
(0.2$<$p$_{\rm _{T}}$$<$0.8~GeV/$c$), and p($\overline{\rm p}$)
(p$_{\rm _{T}}$$<$1.5~GeV/$c$) with a resolution of 5-7\%.

\begin{figure}[t]
\centering
\includegraphics[width=0.40\textwidth]{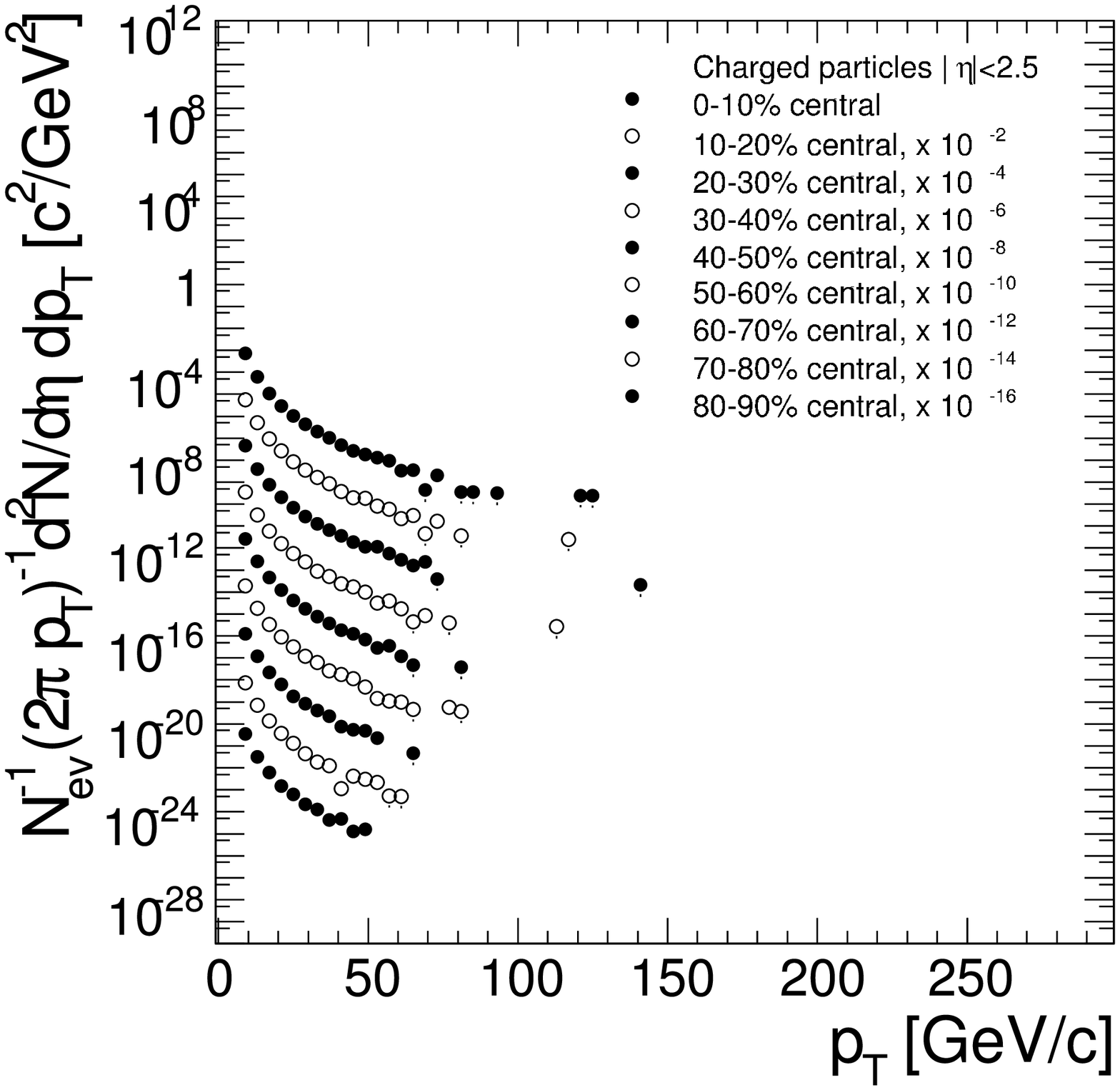}
\includegraphics[width=0.40\textwidth]{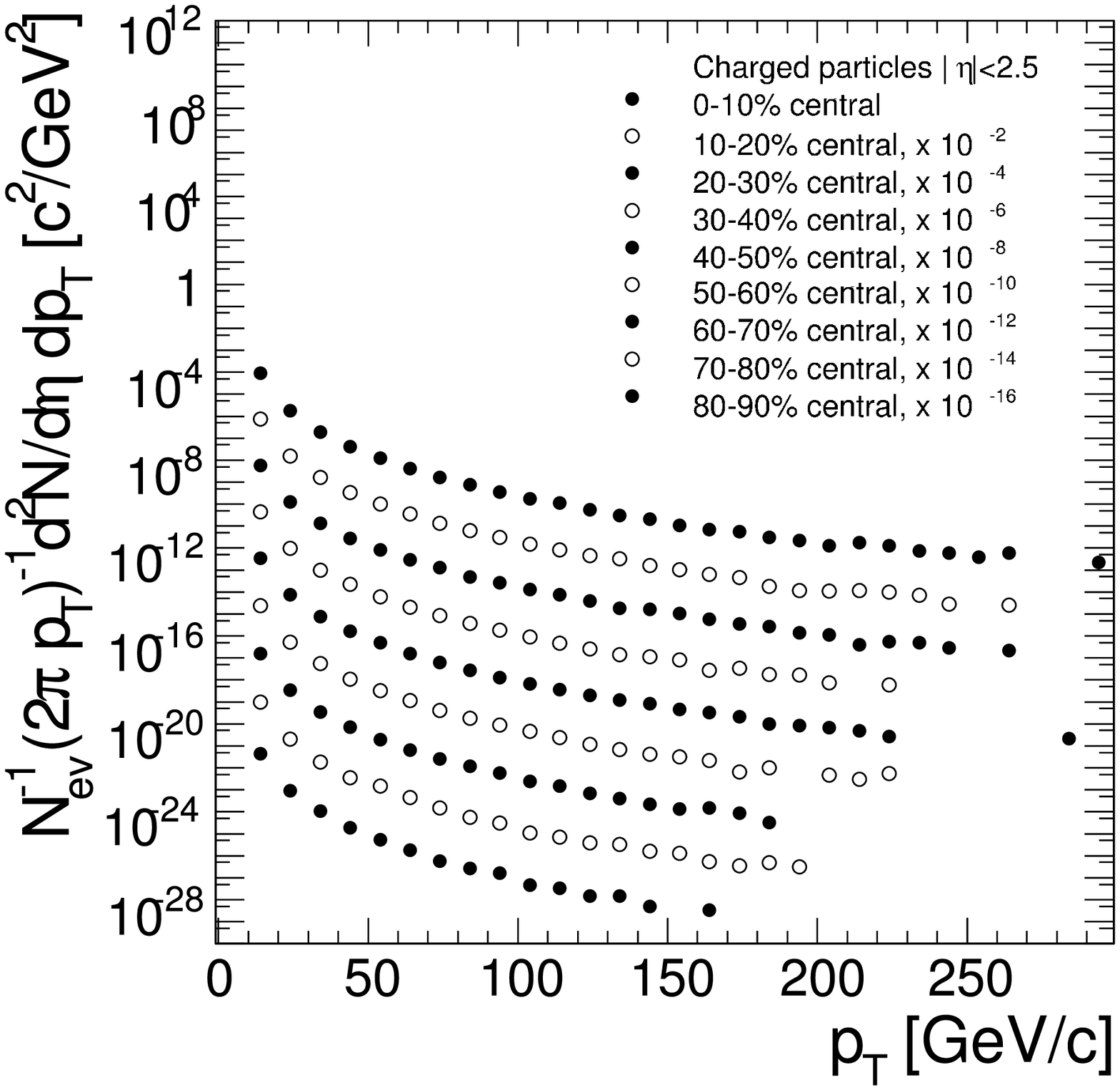}
\caption{\label{fig:pTReach}
Charged particle p$_{\rm _{T}}$ spectra expected for
Pb+Pb collisions at 5.5~TeV with nominal integrated luminosity (0.5
nb$^{-1}$), in 9 centrality bins, offset by factors of 100
for illustration purposes, only using the minimum bias triggered
sample (left) and using the jet-triggered data sets (right).}
\end{figure}

\section{Hard probes}
{\em Hard probes} are of special interest to probe deconfinement properties.
Quarks and gluons, initially created via hard partonic
interactions in the collision, are expected to interact strongly
and lose energy in the resultant hot, dense QCD matter.  Thus,
they are used as a probe of early times in the collision evolution
and a probe of the medium itself.  The suppression of high-p$_{\rm _{T}}$
particles~\cite{cite:PHENIX_highpt_Suppression} (due to medium
induced parton energy loss) and the disappearance of back-to-back
jets in central Au+Au collisions~\cite{cite:STAR_BackToBack} are
two of the main physics results connected with the properties of
the hot, dense QCD medium created at RHIC energies.  Neither of
these effects are observed in cold nuclear matter  ---
peripheral Au+Au or d+Au collisions.

\begin{figure}
\centering
\includegraphics[width=0.40\textwidth]{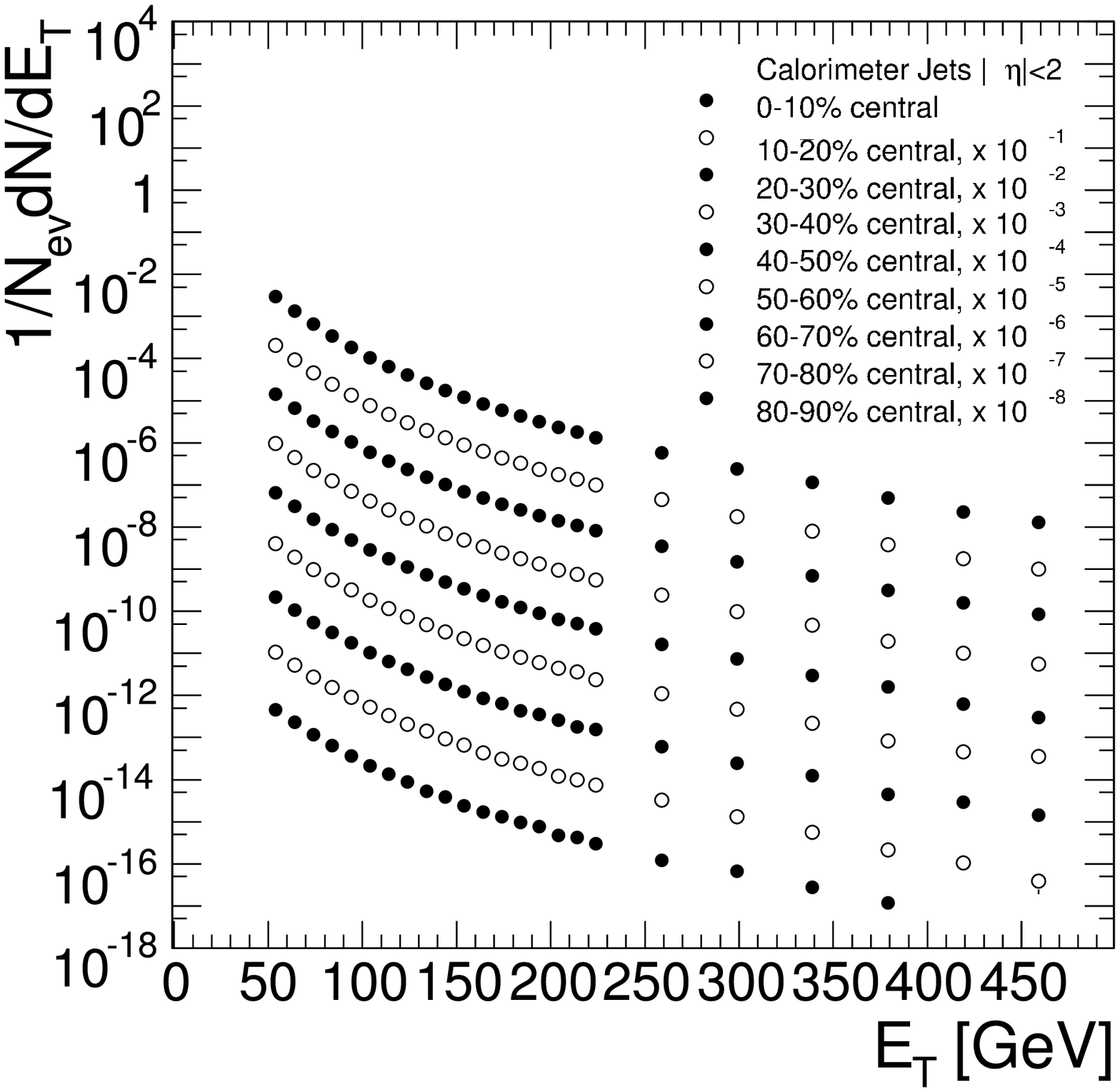}
\includegraphics[width=0.38\textwidth]{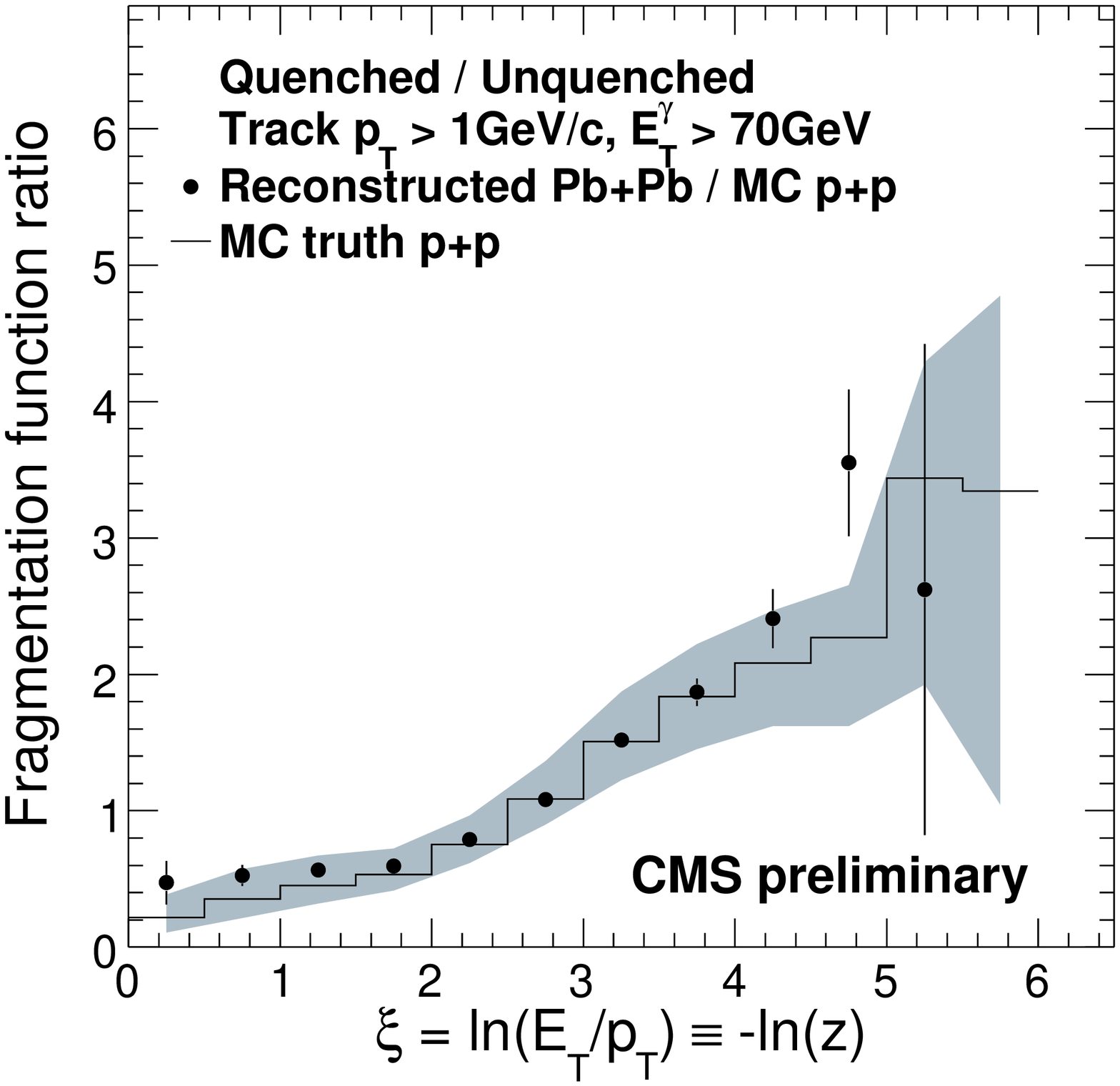}
\caption{\label{fig:Jets}
In the left panel, the expected reach for jets reconstructed by the 
calorimeters.  In the right panel, the ratio of quenched to unquenched
fragmentation functions for $\gamma$-jet reconstruction.}
\end{figure}

The transition from RHIC to the LHC is especially beneficial for
hard probe measurements.  Firstly, the increase in accelerator
luminosities will allow access to more hard scattering events.
Secondly, the vastly increased collision energy results in an
increase in cross-section for hard interactions, that will yield
a large statistical dataset of jets, not obtainable at RHIC.
Thirdly, specific to CMS, a sophisticated high-level triggering
capability will provide an excellent foundation for hard probe
physics. It will extend the p$_{\rm _{T}}$ reach of the collected
events (for charged hadrons and reconstructed jets) by a factor of
three over the number of events recorded in the absence of the trigger
system (see Figs.~\ref{fig:pTReach} and~\ref{fig:Jets}).  
The CMS 4~T magnetic field provides a good
tracking efficiency ($\sim75\%$), resolution ($\sim1.5\%$) and
low fake rate ($<5\%$) for central collision data in the mid-rapidity
barrel region ($\eta<0.5$) for tracks with transverse momenta of up
to 100~GeV/$c$.  Finally, the extensive calorimeter coverage in
both $\eta$ and $\phi$ will provide full jet reconstruction
in heavy-ion collisions.

Jet reconstruction utilises both the hadronic and electromagnetic
calorimeters.  The ``iterative cone'' algorithm with radius R$=$0.5
and background subtraction is used.  An efficiency and purity
of $\sim100\%$ is attained for jets with E$_{\rm _{T}}$$>$$75$~GeV.
The energy resolution for jets with E$_{\rm _{T}}$$>$$100$~GeV is
$\sim15\%$.  In the first year's running (0.5~nb$^{-1}$), the expected
reach for jets is up to a
transverse energy of 0.5~TeV (see left panel of Fig.~\ref{fig:Jets}).

A particular hard scattering interaction that results in back-to-back
$\gamma$-jet production is considered a ``golden'' probe for the study
of in-medium energy loss.  The direct or prompt photon will not
interact with the medium created, thus providing a measure of the
initial jet energy scale (before medium interaction) on an
event-by-event basis.  Isolation of the $\gamma$ is attained by using
combined information from ECAL clusters (photon identification) and HCAL.
Any surrounding jet activity in HCAL is used to veto jets in which
(for example) leading $\pi^{0}(\rightarrow\gamma\gamma$) fragments
also result in ECAL clusters.
To construct the fragmentation function, the ECAL clusters from
isolated photons are correlated with reconstructed calorimeter jets.
A pair of back-to-back $\gamma$-jet is selected if
$\Delta\phi(\gamma$, jet$)$$>$172$^{0}$ is fulfilled.  The E$_{\rm _{T}}$
of the photon is used as an estimate of the 
original E$_{\rm _{T}}$ of the away-side parton.  The jet
fragmentation function ($\xi=ln(\rm E_{_{T}}^{\gamma}/p_{_{T}})$) is
then formed using the charged hadrons reconstructed in the tracker,
on the away-side to the $\gamma$.  Figure~\ref{fig:Jets} (right panel)
shows the ratio of the obtained fragmentation function for the
quenched and unquenched cases.  The expected medium-induced
modification can be discriminated with high significance, where
the high-p$_{\rm _{T}}$ hadrons (low $\xi$) are suppressed.  The jet
energy loss is partially translated into lower momentum particles
(high $\xi$).  Additional details can be found in Ref.~\cite{cite:GammaJetNote}

Heavy quarkonia measurements are {\em the} signature measurements of the
CMS heavy ion program, through reconstructed J/$\psi$ and $\Upsilon$
via di-muons using the muon chambers.  Such measurements will provide
crucial information on the many-body dynamics of high-density QCD matter.
It is generally considered that the step-wise suppression of heavy 
quark-antiquark bound states will be one of the direct probes of
Quark-Gluon Plasma formation~\cite{cite:MatsuiSatz}.  Measuring
quarkonia cross-sections with an
excellent signal-to-background ratio (the best at the LHC) and with a 
mass resolution of 1\% of the quarkonium mass (for both muons in
$|\eta|<2.4$) will allow a {\em clean} study of the J/$\psi$ and
$\Upsilon$ families.  The broad $\eta$ and high-p$_{\rm _{T}}$
acceptance combined with HLT selection will provide a statistical
sample of 1.8$\times$10$^5$ J/$\psi$ and 2.5$\times$10$^4$
$\Upsilon$ from the first year of running (0.5~nb$^{-1}$).

\begin{figure}
\centering
\includegraphics[width=0.45\textwidth]{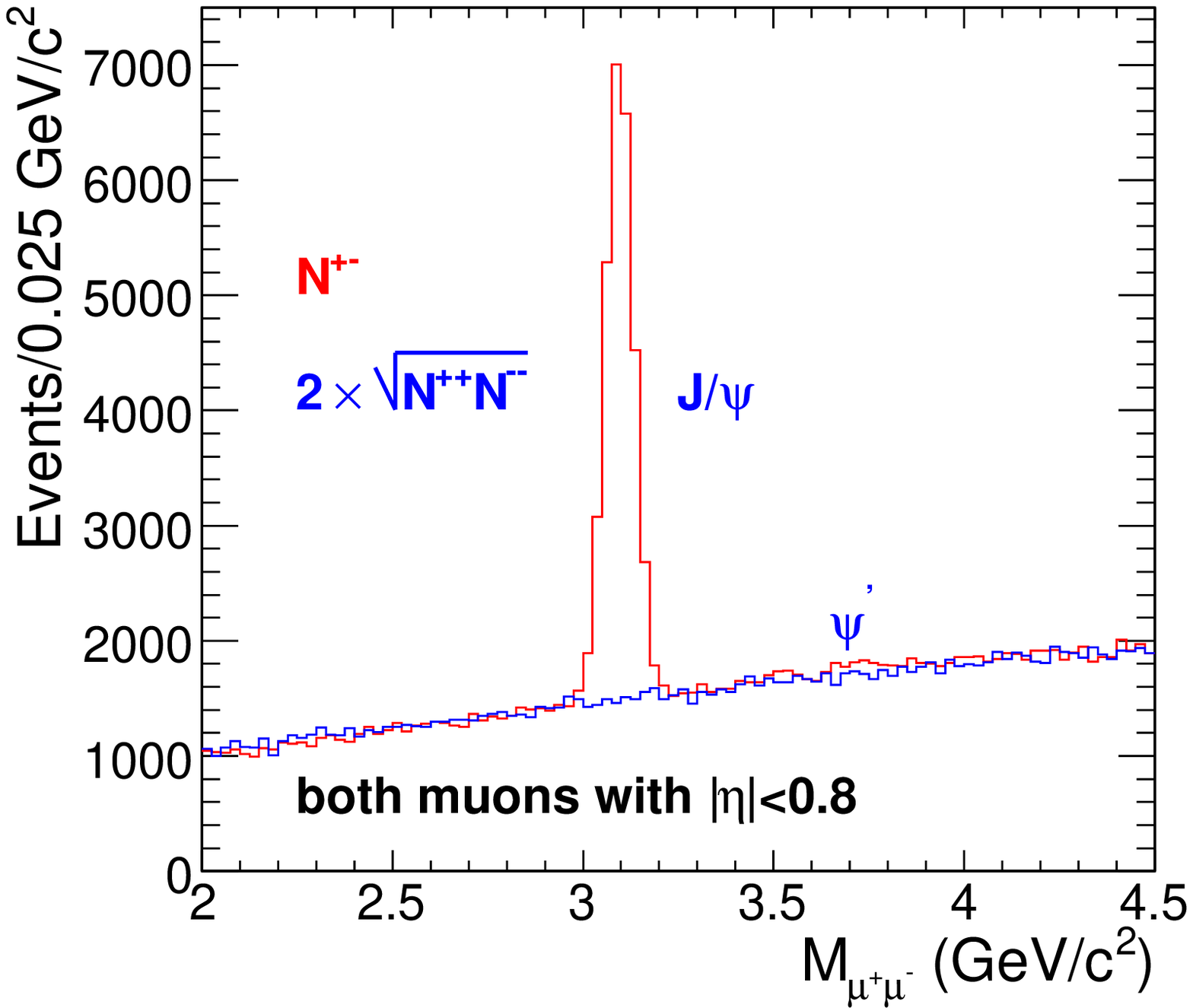}
\includegraphics[width=0.45\textwidth]{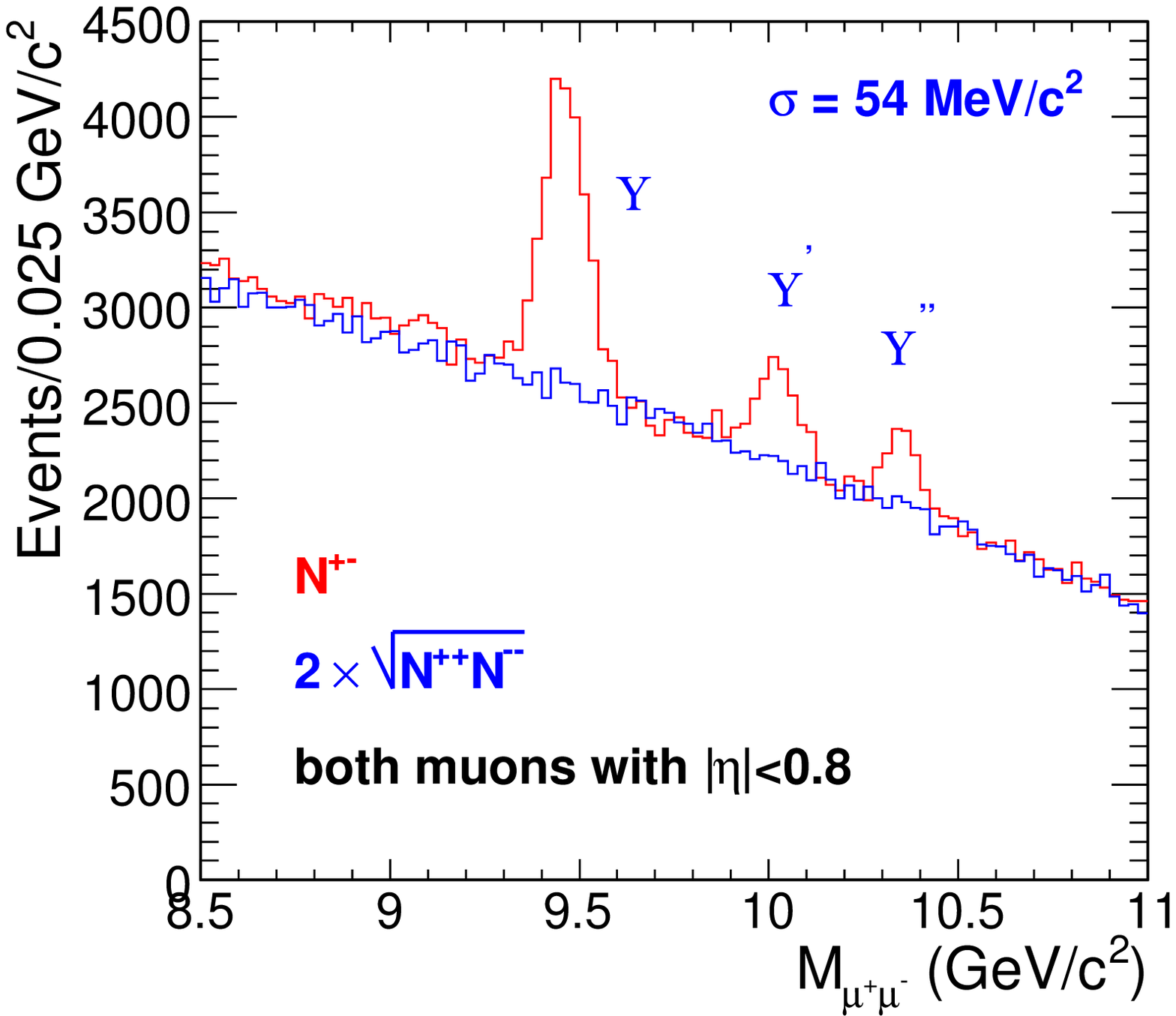}
\caption{\label{fig:HeavyFlavor}
Invariant mass spectra of opposite- and like-sign $\mu$ pairs with 
dN$_{\rm ch}$/d$\eta$$|$$_{\eta=0}$~=~2500 in the J/$\psi$ (left)
and $\Upsilon$ (right) mass regions.  Both $\mu$ have $|\eta|<0.8$ and are 
presented for the expected yield from 1 month of running.}
\end{figure}

\section{Summary}
The design of the CMS detector is suitable to provide a broad range of
physics measurements for the study of high-density QCD.  The highly
segmented Silicon pixel detector is instrumental in the identification
of low momentum particles, in order to study the systematics of bulk
particle production.  The extensive coverage of the calorimeters will
provide measurements of fully reconstructed jets up to
E$_{\rm _{T}}$$\sim$$ 0.5$~TeV in the first year of running
(0.5~nb$^{-1}$).  The muon chambers, coupled with the tracker, will
provide precision measurements of the J/$\psi$ and $\Upsilon$ families
which may shed light on whether a Quark-Gluon Plasma has actually been
formed or not.  A wealth of other important measurements, which include
elliptic flow and ultra peripheral collisions, will lead to a
competitive heavy ion program at LHC.


\section*{References}
\begin{thebibliography}{99}

\bibitem{cite:pTDR} CMS Collaboration, J.~Phys. G34 (2007) 2307.

\bibitem{cite:TDR} CMS Collaboration, CERN/LHCC 94-38 (1994). LHCC/P1.

\bibitem{cite:LowMomTracking} CMS Collaboration, CMS Physics Analysis Summary, QCD-07-001, 2008.

\bibitem{cite:PHENIX_highpt_Suppression} K.~Adcox {\it et al.} Phys.~Rev.~Lett. 88 (2002) 022301.

\bibitem{cite:STAR_BackToBack} C.~Adler {\it et al.} Phys.~Rev.~Lett. 90 (2003) 082302.

\bibitem{cite:GammaJetNote} CMS Collaboration, CMS Physics Analysis Summary, HIN-07-002, 2008.\\ C.~Loizides, {\em proceedings to Quark Matter 2008}, arXiv:0804.3679v1 [nucl-ex]

\bibitem{cite:MatsuiSatz} T.~Matsui and H.~Satz, Phys. Lett {\bf B178} (1986) 416

\end{thebibliography}
\end{document}